\documentclass[prl,aps,twocolumn,10pt,superscriptaddress]{revtex4-2} 
\usepackage{amsmath,graphicx,color,gensymb,bm}
\usepackage{dcolumn}
\usepackage{amsthm}     
\usepackage{amsfonts}
\usepackage{algorithmic}
\usepackage{enumerate}
\usepackage{latexsym}
\usepackage{amssymb}
\usepackage{multirow}
\usepackage{array}
\usepackage{diagbox}
\usepackage{subfigure}
\usepackage[colorlinks=true,citecolor=blue,linkcolor=blue]{hyperref}

\begin{document}

\preprint{APS/123-QED}

\title{Time-reversal symmetry breaking superconductivity in the presence of loop-current fluctuations}
\author{Zenghui Fan}
\affiliation{School of Physics and Astronomy, Beijing Normal University, and Key Laboratory of Multiscale Spin Physics (Beijing Normal University), Ministry of Education, Beijing 100875, China}
\author{Runyu Ma}
\affiliation{School of Physics and Astronomy, Beijing Normal University, and Key Laboratory of Multiscale Spin Physics (Beijing Normal University), Ministry of Education, Beijing 100875, China}

\author{Stefano Chesi}
\affiliation{Beijing Computational Science Research Center, Beijing 100084, China\\}

\author{Congjun Wu}
\email{wucongjun@westlake.edu.cn}
\affiliation{New Cornerstone Science Laboratory, Department of Physics, School of Science, Westlake University, Hangzhou 310024, Zhejiang, China}
\affiliation{Institute for Theoretical Sciences, Westlake University, Hangzhou 310024, Zhejiang, China}
\affiliation{Key Laboratory for Quantum Materials of Zhejiang Province, School of Science, Westlake University, Hangzhou 310024, Zhejiang, China}
\affiliation{Institute of Natural Sciences, Westlake Institute for Advanced Study, Hangzhou 310024, Zhejiang, China}

\author{Tianxing Ma}
\email{txma@bnu.edu.cn}
\affiliation{School of Physics and Astronomy, Beijing Normal University, and Key Laboratory of Multiscale Spin Physics (Beijing Normal University), Ministry of Education, Beijing 100875, China}

\begin{abstract}
Loop currents have been proposed in various superconductors and recently confirmed in kagome materials, raising a fundamental question regarding their intrinsic connection to superconductivity. Here, we study a sign-problem-free bilayer $t-J_{\perp}-V$ model hosting a spontaneous interlayer loop-current parent state, and explore the interplay between loop-current fluctuations and superconductivity using unbiased projector quantum Monte Carlo simulations. Near half-filling, unbiased interlayer interactions induce spontaneous loop currents that break time-reversal symmetry. Upon hole doping, the loop-current order is suppressed, and interlayer $s$-wave superconductivity emerges where loop-current fluctuations become dominant. We establish a phase diagram revealing a transition from the loop-current parent to a superconducting state, reminiscent of the evolution from an antiferromagnetic parent to superconductivity in cuprates. Strikingly, a coexisting regime emerges near the phase boundary, yielding time-reversal-symmetry-breaking superconductivity. Our study reveals an intrinsic connection between loop currents and superconductivity, and identifies a promising mechanism for time-reversal symmetry breaking in superconductors. Furthermore, our results offer insights into unconventional superconductivity in loop-current systems and establish a minimal theoretical framework for understanding time-reversal symmetry breaking in bilayer correlated electron systems.
\end{abstract}
\maketitle

\noindent
\underline{\it Introduction}~~
The discovery of copper-based superconductors~\cite{WOS:A1986D856200010} established high-temperature superconductivity (SC) as a central topic in condensed matter physics.
Clarifying its mechanism is essential for discovering new high-$T_c$ materials.
In many kinds of superconductors~\cite{WOS:A1986D856200010,WOS:000253951900032,WOS:A1980JG10700005,WOS:001052549700001}, SC emerges alongside various fluctuations, which are often pivotal to its formation~\cite{WOS:000258326000009,PhysRevLett.109.187004,WOS:001437303700022}.
For instance, in doped cuprates, antiferromagnetic (AFM) fluctuations have been closely tied to spin-singlet pairing near a quantum critical point (QCP)~\cite{WOS:000349190300029,yuan2022scaling}.
Other fluctuations, including charge density waves~\cite{WOS:000368015700031}, spin density waves~\cite{ZHAO20251239}, and stripe order~\cite{doi:10.1126/science.aam7127}, also play crucial roles, resulting in rich phase diagrams.
Recently, significant attention has been drawn to orbital magnetic fluctuations that break time-reversal symmetry (TRS) in kagome superconductors~\cite{WOS:000753550800017,WOS:000866362700006,xing2024optical,asaba2024evidence,FENG20211384,LIU20251211,PhysRevLett.134.096401,cpl_41_4_047103}.
This orbital magnetism is associated with a pattern of spontaneously flowing currents.
According to a theory by Bloch~\cite{PhysRev.75.502,doi:10.1143/JPSJ.65.3254,PhysRevResearch.4.013043}, such currents must form closed loops, known as loop currents (LCs), to avoid forbidden global currents.
Yet, the intrinsic connection between LC fluctuations and SC remains a fundamental open question.

In strongly correlated systems, magnetic fluctuations arise not only from spin-related interactions but also in the orbital sector through LCs.
The identification of such currents in diverse superconducting systems---including cuprates~\cite{PhysRevB.55.14554,PhysRevB.67.054511,PhysRevB.73.155113}, twisted bilayer graphene~\cite{PhysRevB.87.035427}, and kagome systems~\cite{WOS:000753550800017,WOS:000866362700006,xing2024optical,asaba2024evidence,FENG20211384,LIU20251211,PhysRevLett.134.096401,cpl_41_4_047103}---has made their relationship with SC a critical and actively debated issue.

In high-$T_c$ cuprates, LC order has been proposed as a hidden order underlying the pseudogap phase~\cite{PhysRevB.55.14554,PhysRevB.67.054511,PhysRevB.73.155113}.
Its possible connection to $d_{x^2-y^2}$-wave superconducting pairing near the quantum critical point has been extensively discussed~\cite{PhysRevB.81.064515}.
Furthermore, LC order has been invoked to interpret polarized neutron scattering experiments~\cite{PhysRevLett.96.197001,li2008unusual,li2010hidden,bourges2021loop}.
However, the absence of LC signatures in other studies~\cite{PhysRevLett.111.187003,wu2015incipient,PhysRevB.101.184511,PhysRevB.103.134426} has led to persistent controversy regarding the interplay between LC fluctuations and SC.
Recently, the observation of a TRS-breaking charge density wave in kagome superconductors AV$_3$Sb$_5$ (A = K, Rb, Cs)~\cite{WOS:000753550800017,WOS:000866362700006,xing2024optical,asaba2024evidence,FENG20211384,LIU20251211} has revitalized the LC proposal and intensified interest in their relationship with SC~\cite{palle2024sc,cui2026minimal,PhysRevLett.131.226001,Li_2025,10.1093/nsr/nwaf414,5vyy-rj6v}.
Theoretical models attribute these currents to nonlocal interactions~\cite{10.1093/nsr/nwaf414,5vyy-rj6v,cpl_42_8_080709}, yet an unbiased, numerically exact characterization of the LC-SC interplay remains elusive due to the fermion sign problem.
One promising strategy to circumvent this challenge is the interlayer loop-current (ILC) framework.
In prior work by one of the authors~\cite{PhysRevB.70.220505}, this configuration was termed interlayer staggered currents, providing a foundation for our present study.

To address this challenge, we introduce a minimal bilayer model in which spontaneous ILCs emerge from unbiased interlayer exchange and Coulomb repulsive interactions, free from the fermion sign problem.
LC models typically involve directly incorporating a biased weighting of the LC pattern into the Hamiltonian.
By contrast, our framework avoids such presumptions, allowing an intrinsic ILC parent to develop microscopically.
The resulting bilayer structure and significant interlayer exchange coupling are shared features with a broader class of correlated electron systems, including the recently discovered bilayer nickelate La$_3$Ni$_2$O$_7$~\cite{XIE20243221,PhysRevLett.132.146002,WOS:001052549700001}.
However, our primary aim is not to quantitatively describe a specific material, but to establish a controlled theoretical platform for investigating the interplay between ILC fluctuations and SC in layered correlated systems.

\begin{figure}[tb]
\includegraphics[scale=0.156]{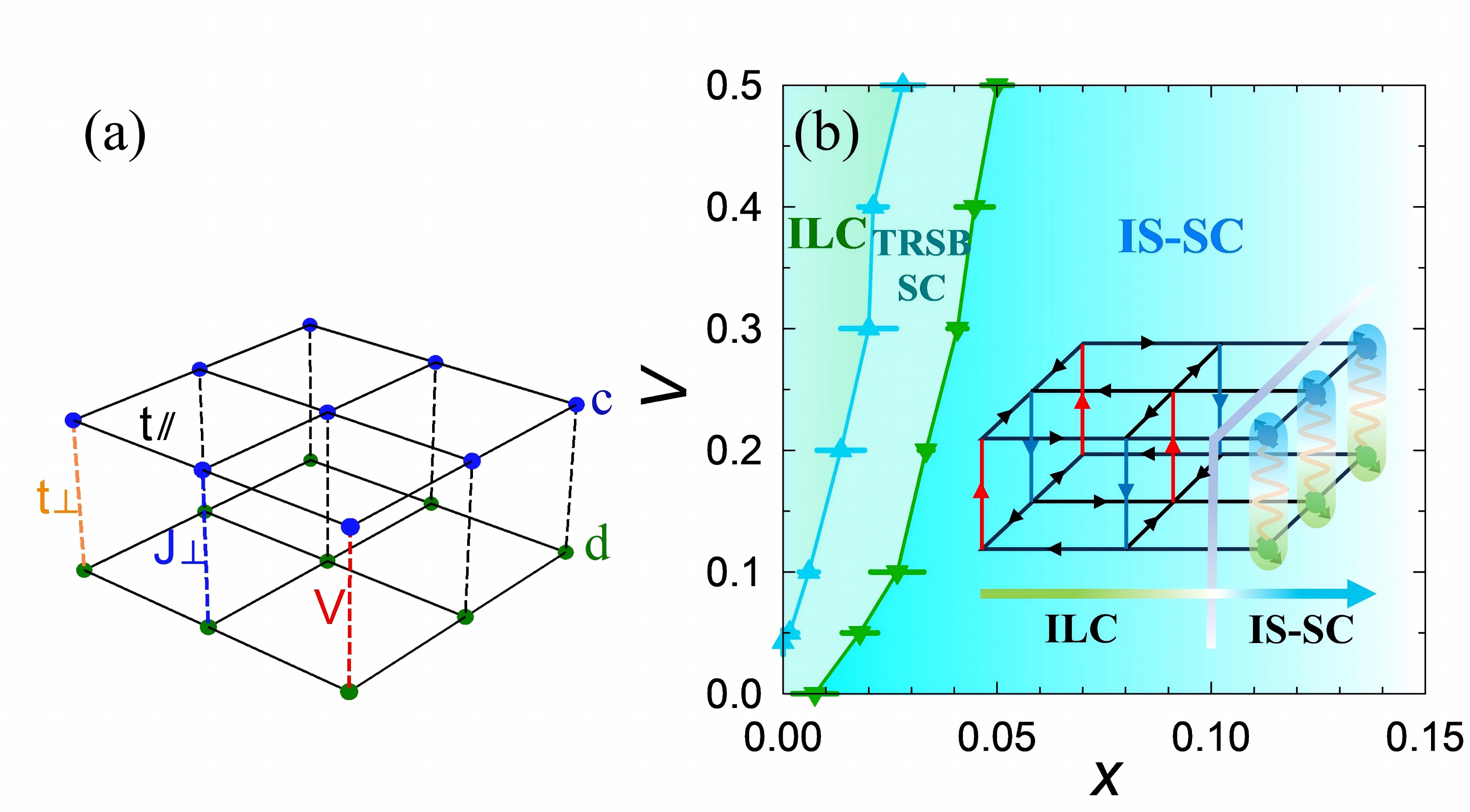}
\caption{\label{Fig_1} (a) Sketch of the bilayer model with intralayer nearest-neighbor hopping $t_{\parallel}$, interlayer on-site hopping $t_{\perp}$, interlayer antiferromagnetic spin-exchange coupling $J_{\perp}$, and interlayer Coulomb repulsion $V$. (b) Phase diagram as a function of hole doping $x$ and interlayer Coulomb repulsion $V$ at $J_{\perp}=2$, showing the interlayer loop-current (ILC) phase, the interlayer $s$-wave superconducting (IS-SC) phase, and a coexisting phase with significant time-reversal-symmetry-breaking superconductivity (TRSB SC). The green and blue solid lines mark the phase boundaries of ILC and IS-SC, respectively, as determined by finite-size scaling. Inset: schematic illustrations of the ILC and IS-SC patterns.}
\end{figure}

In this work, we investigate a $t$-$J_{\perp}$-$V$ model on a hole-doped bilayer square lattice, as sketched in Fig.~\ref{Fig_1}(a), using the unbiased projector quantum Monte Carlo (PQMC) method~\cite{Assaad2008,LI2016925}.
This model is free from the sign problem due to Kramers invariance~\cite{PhysRevLett.91.186402}, thereby overcoming the difficulties of simulating doped systems and yielding numerically exact results.
Our calculations demonstrate that spontaneous ILCs are stabilized by nearest-neighbor interlayer interactions near half-filling, forming an ILC parent state analogous to the proposed nonlocal physics in kagome models~\cite{10.1093/nsr/nwaf414,5vyy-rj6v}.
Upon hole doping, the ILC order is rapidly suppressed, while interlayer $s$-wave superconductivity (IS-SC) emerges with an optimal doping behavior, reminiscent of the evolution from an antiferromagnetic parent to superconductivity.
Structure factors and correlation lengths reveal a strong interplay between ILC fluctuations and IS-SC, with superconducting correlations peaking where ILC fluctuations are pronounced.
The resulting phase diagram, summarized in Fig.~\ref{Fig_1}(b), highlights a transition from the ILC state to SC, with a coexisting phase emerging near the quantum critical point.
This coexistence is characterized by SC with significant TRS breaking, reflecting the persistence of LC order within the superconducting phase.
Our model directly addresses the pivotal question regarding the intrinsic connection between LCs and SC, establishing that IS-SC emerges in the presence of strong ILC fluctuations.
Furthermore, we uncover a mechanism for TRS breaking in superconductivity, offering insights into related phenomena in layered correlated systems, including recent observations in bilayer nickelates~\cite{ji2026timereversal}.

\noindent
\underline{\it Results and discussion}~~
LCs are thought to bear an important connection to SC, as highlighted by proposals in high-$T_c$ cuprates~\cite{PhysRevB.81.064515,PhysRevLett.102.017005}.
Recent discussions of interlayer $s$-wave pairing in bilayer systems have emphasized the importance of strong interlayer exchange coupling~\cite{PhysRevLett.132.146002}.
Motivated by these considerations, our bilayer model allows us to directly investigate the intrinsic connection between LCs and SC.
The resulting phase diagram [Fig.~\ref{Fig_1}(b)] reveals a transition from a TRS-breaking ILC parent to a superconducting state.
Near half-filling, spontaneous ILCs are stabilized by interlayer exchange $J_{\perp}=2$ and further strengthened by interlayer Coulomb repulsion $V$, expanding the ILC region.
With increasing hole doping, the ILC order is rapidly suppressed, while IS-SC emerges.
Notably, the superconducting correlations are established where ILC fluctuations are pronounced, indicating an intimate interplay between the suppression of LC order and the emergence of SC.
This evolution is reminiscent of the transition from an antiferromagnetic parent to superconductivity in cuprates, with the distinction that loop currents contribute orbital magnetism rather than spin magnetism.

Recently, experimental observations have reported an ``unprecedented superconducting state'' in bilayer nickelate superconductors~\cite{ji2026timereversal}---a TRS-breaking superconducting state emerging in the low-temperature regime below the superconducting transition and extending even to the ground state.
Remarkably, the ILCs in our model break time-reversal symmetry, and our ground-state phase diagram reveals a coexistence region of ILC and IS-SC phases, in which SC inherently exhibits significant TRS breaking.
Even within the predominantly superconducting region, signatures of TRS breaking may persist, reflecting the influence of residual loop-current fluctuations.
This coexistence establishes a mechanism for TRS breaking in SC.
While our minimal model is not intended to quantitatively describe a specific material, its bilayer geometry and dominant interlayer coupling capture essential ingredients shared by layered correlated systems, offering a theoretical framework for understanding TRS breaking in such platforms.

\begin{figure}[tb]
\includegraphics[scale=0.355]{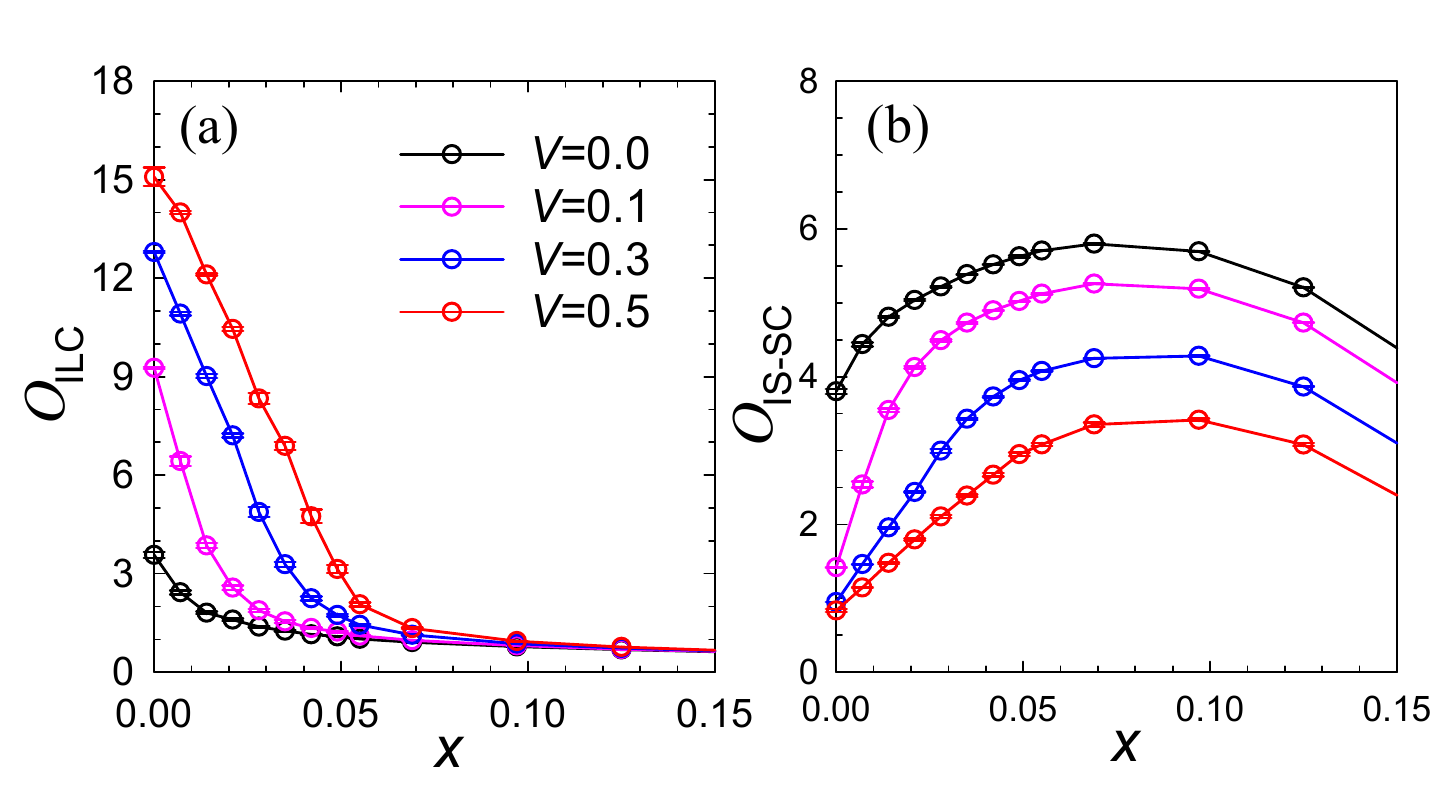}
\caption{\label{Fig_2} Structure factors (a) $O_{\mathrm{ILC}}$ and (b) $O_{\mathrm{IS-SC}}$ as functions of hole doping $x$ for different interlayer Coulomb repulsions $V$ . The ILC structure factor is suppressed by $x$ and significantly enhanced by $V$ near half-filling, while the IS-SC structure factor is suppressed by $V$ and exhibits optimal doping behavior.}
\end{figure}

We first examine the structure factors of the interlayer current and the interlayer $s$-wave pairing (Eqs.~\ref{uniform} and \ref{sc}).
The interlayer pairing and current structure factors peak at wavevectors $(0,0)$ and $(\pi,\pi)$, respectively.
The latter indicates a staggered ILC with broken TRS in our sign-problem-free model [please see {\color{blue} Supplemental Material (SM), Sec.~S1} for detailed evidence].
Fig.~\ref{Fig_2} displays the ILC and IS-SC structure factors as functions of doping $x$ for different interlayer Coulomb repulsions $V$ at a fixed system size $L=12$.
The ILC structure factor decreases rapidly with doping but is significantly enhanced by $V$ near half-filling.
In contrast, the IS-SC structure factor is suppressed by $V$ and enhanced by doping with an optimal doping behavior.
Notably, the maximum of the IS-SC structure factor roughly coincides with the suppression of the ILC signal, indicating an intimate interplay between the suppression of LC order and the emergence of SC.
The opposing effects of $V$ can be physically understood: $V$ breaks the charge occupation balance between the two layers, thereby enhancing interlayer charge transfer while suppressing uniform interlayer charge pairing.
Analogously, we also consider the AFM structure factor, which is likewise suppressed by $V$ [please see {\color{blue}SM, Sec.~S3} for details].
The role of $V$ is analogous to the nonlocal interaction proposed for LCs in kagome models~\cite{10.1093/nsr/nwaf414,5vyy-rj6v}.
These results further reveal an intriguing competition between ILC order and IS-SC in the bilayer system.

\begin{figure}[tb]
\includegraphics[scale=0.36]{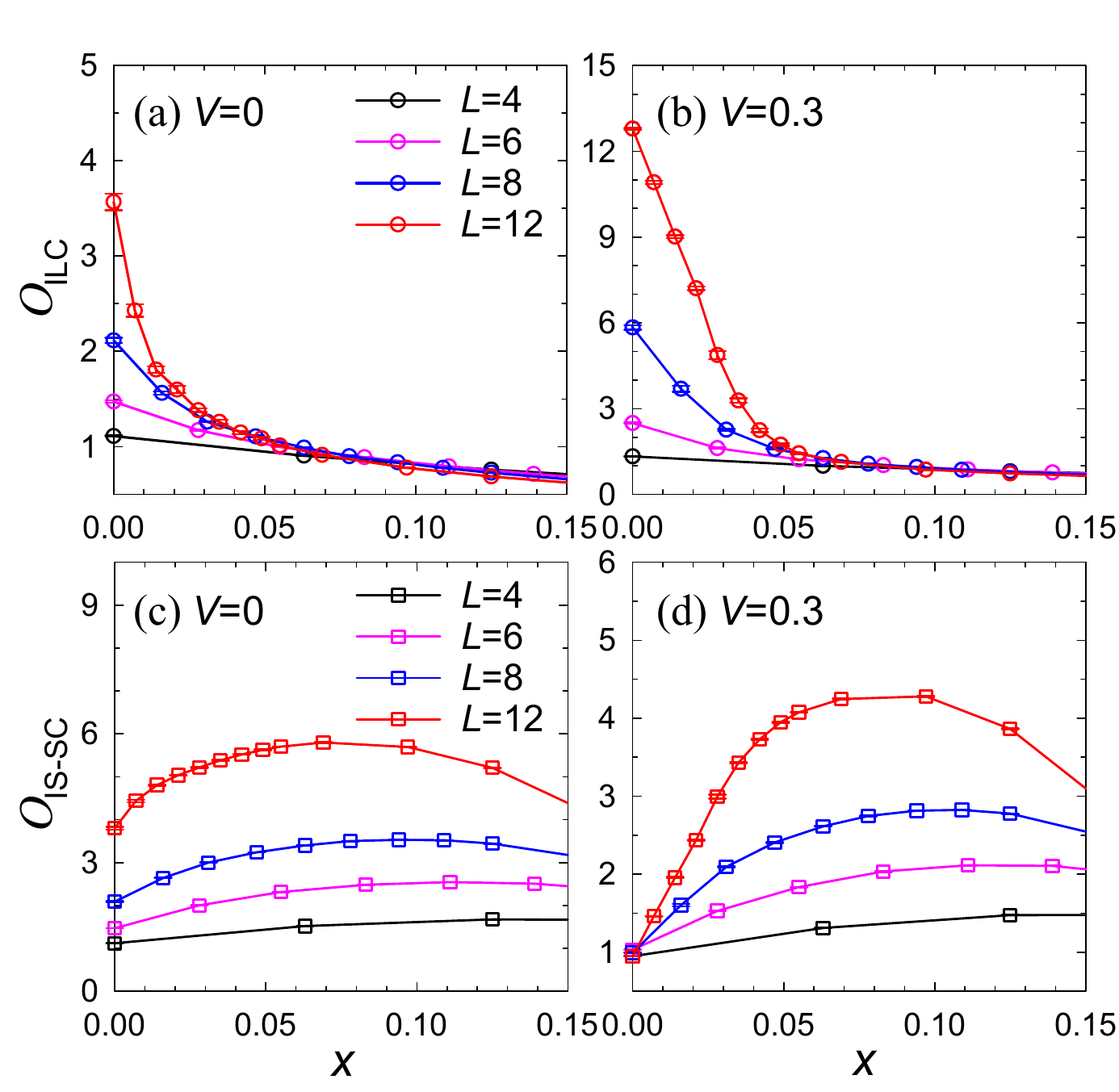}
\caption{\label{Fig_3} Structure factors $O_{\mathrm{ILC}}$ and $O_{\mathrm{IS-SC}}$ as functions of doping $x$ with different lattice size $L$ at (a) $V=0$ and (b) $V=0.3$ for ILC, and at (c) $V=0$ and (d) $V=0.3$ for IS-SC. The $L$-dependence of structure factor is revealed in this figure, where the growth with increasing $L$ indicates the trend for ordering, while the saturation indicates disorder. 
}
\end{figure}

To assess the presence of long-range order, we examine the system size $L$-dependence of the structure factors, which qualitatively reflects the behavior of the correlation length $\xi$, as shown in Fig.~\ref{Fig_3}.
In a disordered phase (i.e., $\xi\ll L$), the structure factor becomes independent of $L$ and saturates to a value proportional to $\xi^2$.
Conversely, in an ordered phase (i.e., $\xi\gg L$), it is limited by the system size and grows with increasing $L$.
For the ILC structure factor in Figs.~\ref{Fig_3}(a) and (b), the $L$-dependence is suppressed by doping $x$ but enhanced by the interlayer repulsion $V$ near half-filling.
In contrast, the IS-SC exhibits significant $L$-dependence over a broad doping range (as shown in Figs.~\ref{Fig_3}(c) and (d)), which is suppressed by $V$ near half-filling (e.g., especially for $V=0.3$ in panel (d)).
Notably, the most pronounced $L$-dependence for the IS-SC occurs approximately where the ILC signal is suppressed, highlighting the competition between these two phases and suggesting that SC emerges where LC fluctuations become dominant.

\begin{figure}[t]
\includegraphics[scale=0.14]{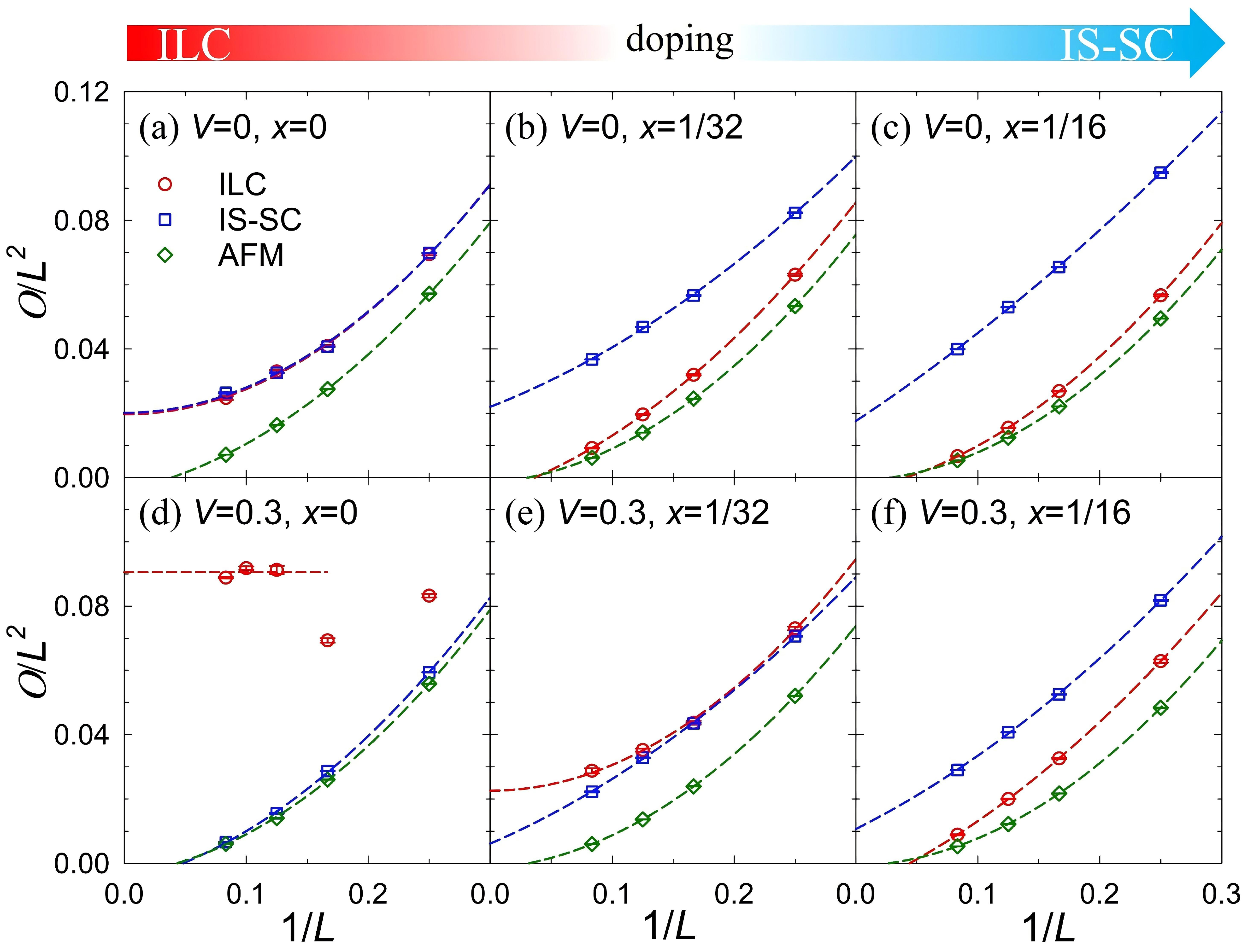}
\caption{\label{Fig_4} Structure factors $O/L^{2}$ as functions of the reciprocal of lattice size $1/L$ for ILC, IS-SC and AFM. The data with different sizes are extrapolated to the thermodynamic limit at different doping $x=0$, $x=1/32$ and $x=1/16$ for the interlayer repulsion (a)-(c) $V=0$ and (d)-(f) $V=0.3$ respectively. The finite-size scaling is performed by the polynomial fittings to the data of $L=4, 6, 8, 12$ denoted by dashed lines. A special instruction is that a additional $L=10$ for ILC is added at (d) and the red flat dashed line of (d) is a guide for eyes.}
\end{figure}

\begin{figure}[tb]
\includegraphics[scale=0.37]{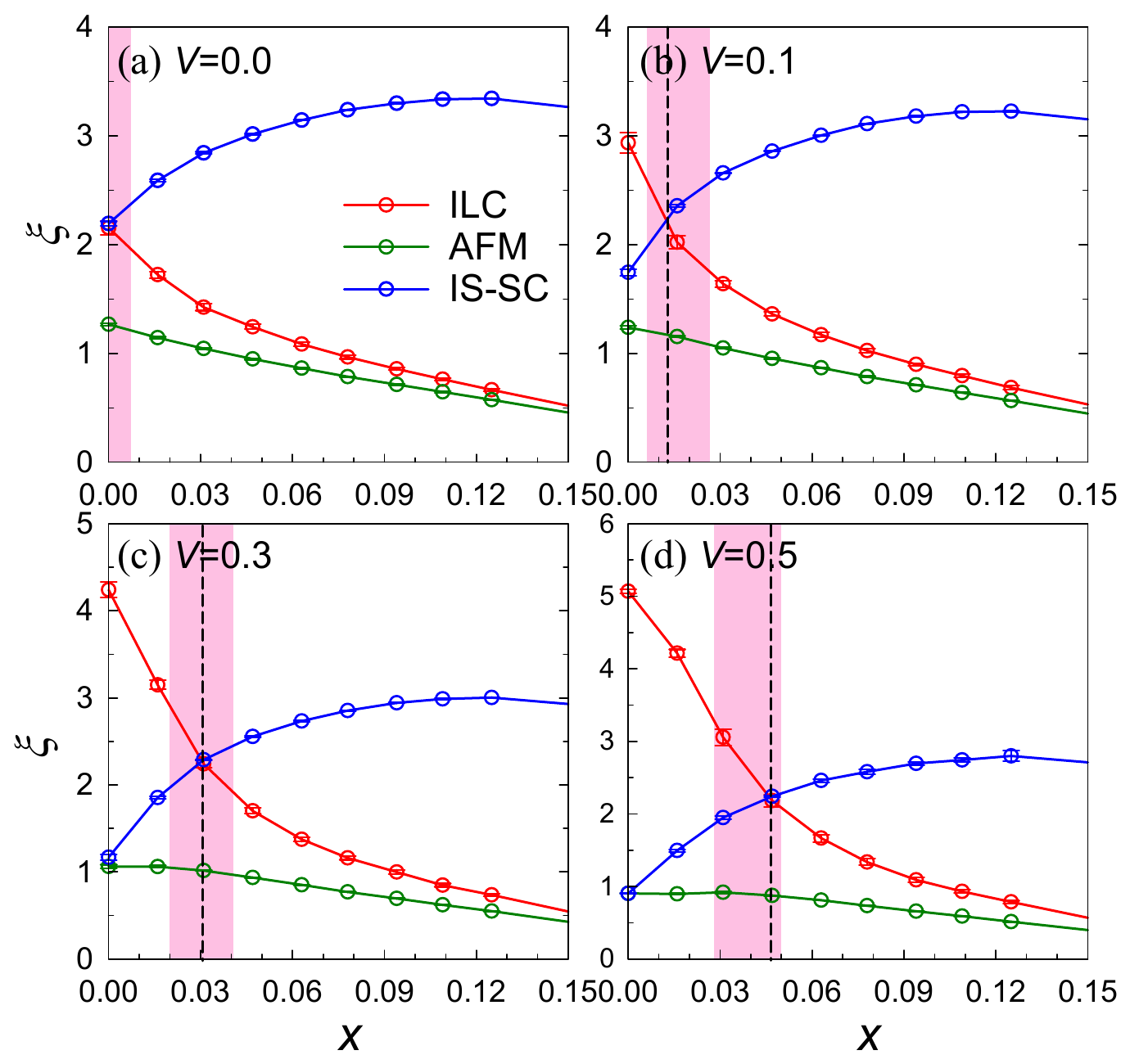}
\caption{\label{Fig_5} Correlation lengths $\xi$ as functions of doping $x$ for ILC, AFM and IS-SC with different interlayer repulsion (a) $V=0.0$, (b) $V=0.1$, (c) $V=0.3$ and (d) $V=0.5$ at $L=8$. The crossover where the dominant order changes is marked by a black dashed line. These crossovers consistently fall into the pink areas which record the region of coexisting phase as shown in Fig.~\ref{Fig_1}.
}
\end{figure}

To further determine the phase boundaries, we perform a finite-size scaling analysis of the structure factors to extract their values in the thermodynamic limit, as shown in Fig.~\ref{Fig_4}, based on the scaling hypothesis $O_{\alpha}(L)/L^{2}=a\xi^{2}/L^{2}+b/L+c$~\cite{PhysRevX.3.031010}.
We first consider the case with no interlayer repulsion, $V=0$, as shown in Figs.~\ref{Fig_4}(a)--(c).
At half-filling ($x=0$), we find that ILC establishes long-range order in the thermodynamic limit, as indicated by a finite intercept at $L\rightarrow\infty$.
This demonstrates the qualitative role of interlayer exchange $J_{\perp}$ in stabilizing ILCs in the absence of $V$.
Additionally, IS-SC similarly exhibits long-range order, indicating a coexisting phase at this parameter set.
With increased doping to $x=1/32$ and $x=1/16$, the IS-SC long-range order persists, while the ILC order is suppressed.
Introducing a finite interlayer repulsion $V=0.3$ alters this picture, as shown in Figs.~\ref{Fig_4}(d)--(f).
At half-filling, the ILC order is strongly enhanced and robust, while IS-SC order is absent.
Notably, in panel (d), the upward curvature of the $O_{\mathrm{ILC}}/L^2$ data at $L=6$ strongly indicates a pronounced tendency toward ILC ordering.
To ensure reliability, the inclusion of $L=10$ data in panel (d) confirms this trend, showing a large $O_{\mathrm{ILC}}/L^2$ approaching saturation at large $L$.
At $x=1/32$, both ILC and IS-SC exhibit long-range order, but by $x=1/16$, the ILC order vanishes, leaving only the IS-SC long-range order.
These results clearly display a transition from an ILC parent to a doped IS-SC phase, resembling the conventional AFM--SC transition.
Strikingly, the concomitant long-range orders for ILC and IS-SC at different $V$ and $x$ indicate a parameter-dependent coexisting phase near the quantum critical point, as shown in Fig.~\ref{Fig_1}(b), where SC exhibits significant TRS breaking due to the persistence of LC order.
We also note that antiferromagnetic order is absent for all parameters in our system, as shown in all panels of Fig.~\ref{Fig_4}.
An interpolation method was used to obtain data at fixed dopings for different system sizes.
The phase boundaries for ILC and IS-SC are determined from this detailed finite-size scaling and summarized in the phase diagram of Fig.~\ref{Fig_1}(b).

The correlation length $\xi$ defined in Eq.~\ref{length} directly reflects the competition among different orders, as illustrated in Fig.~\ref{Fig_5}.
Across all parameters, the AFM correlation length is always exceeded by that of ILC and IS-SC, consistent with the finite-size scaling analysis presented above.
A key observation is the opposing evolution of the ILC and IS-SC correlation lengths upon doping: the ILC correlation length declines while the IS-SC correlation length grows, indicating strong competition between these two phases and suggesting that SC emerges where LC fluctuations are pronounced.
The crossover point of these curves in Fig.~\ref{Fig_5} marks the crossover between dominant phases.
As $V$ increases, this critical doping shifts to higher values, and the transition occurs within the narrow coexistence region of the phase diagram, highlighted in pink in Fig.~\ref{Fig_5}.
Furthermore, the correlation length is also used to confirm the dominance of the interlayer $s$-wave pairing in the SC channel for our bilayer model in {\color{blue}SM, Sec.~S4}.

\noindent
\underline{\it Conclusions}~~
Centering on the recently revitalized topic of LCs, we directly address a fundamental question in condensed matter physics---the intrinsic connection between LC fluctuations and SC---through a minimal quantum system hosting a spontaneous LC parent state. Using unbiased PQMC simulations, we investigate a microscopic $t$-$J_{\perp}$-$V$ model that offers three key advantages: (1) The model spontaneously generates ILCs near half-filling without biasing the Hamiltonian, thereby breaking TRS. (2) Owing to Kramers invariance, the model is free from the fermion sign problem, enabling numerically exact results for doped systems. (3) Its bilayer geometry and significant interlayer exchange coupling capture essential ingredients shared by a broader class of layered correlated electron systems, including the recently discovered bilayer nickelates.

Our results demonstrate that ILCs are spontaneously established by interlayer exchange $J_{\perp}$ and further stabilized by interlayer repulsion $V$ near half-filling, forming an intrinsic ILC parent state. Upon hole doping, the ILC order is rapidly suppressed, while IS-SC emerges with an optimal doping profile. Notably, the enhancement of SC correlations coincide with the suppression of LC order, indicating competitive behavior between the two phases and suggesting that SC emerges in the regime where LC fluctuations are dominant. Based on long-range order analysis, we construct a phase diagram [Fig.~\ref{Fig_1}(b)] revealing a transition from the ILC parent to a superconducting state. This evolution is reminiscent of the transition from an AFM parent to SC in cuprates, except that LCs contribute orbital rather than spin magnetism. Strikingly, a coexisting region emerges near the quantum critical point, characterized by SC with significant TRS breaking, reflecting the persistence of LC order within the superconducting phase. This coexistence establishes a mechanism for TRS breaking in SC.

In conclusion, our model directly addresses the pivotal question regarding the intrinsic connection between LCs and SC, establishing that IS-SC emerges in the presence of strong LC fluctuations. Furthermore, we uncover a mechanism for TRS breaking in SC, offering insights into related phenomena in layered correlated systems. Moreover, our results provide important insights into unconventional SC in LC systems.

Our study may also connects to experimental advances in ultracold atoms. Mixed-dimensional bilayer optical lattices with strong interlayer spin-exchange interactions have recently been realized~\cite{WOS:000955590300007,WOS:000787150100001}. Furthermore, related $t$-$J$ models with rung Coulomb repulsion have also been proposed for realization in ultracold atom systems~\cite{PhysRevB.109.045127,PhysRevB.110.L081113}. These developments establish ultracold atom platforms as promising candidates for realizing our minimal model and probing the interplay between LC fluctuations and SC in a controlled setting.

\noindent
\underline{\it Model and methods}~~
The $t$--$J_{\perp}$--$V$ Hamiltonian studied in this work is defined as
\begin{eqnarray}
\label{Hamiltonian}
H&=&-t_{\parallel}\sum_{\langle{\bf ij}\rangle\sigma}(c_{{\bf i}\sigma}^\dagger c_{{\bf j}\sigma}
+d_{{\bf i}\sigma}^\dagger d_{{\bf j}\sigma} + \mathrm{H.c.}) \nonumber\\
&&-t_{\perp}\sum_{{\bf i}\sigma}(c_{{\bf i}\sigma}^\dagger d_{{\bf i}\sigma} + \mathrm{H.c.}) 
-\mu\sum_{{\bf i}\sigma}(c_{{\bf i}\sigma}^\dagger c_{{\bf i}\sigma}
+d_{{\bf i}\sigma}^\dagger d_{{\bf i}\sigma})  \nonumber\\
&&+J_{\perp}\sum_{\bf i}\bm{S}_{{\bf i},c}\cdot \bm{S}_{{\bf i},d}+V\sum_{\bf i}(n_{{\bf i},c}-1)(n_{{\bf i},d}-1),
\end{eqnarray}
where $c_{{\bf i}\sigma}$ ($c_{{\bf i}\sigma}^\dagger$) and $d_{{\bf i}\sigma}$ ($d_{{\bf i}\sigma}^\dagger$) are electron annihilation (creation) operators with spin $\sigma=\uparrow,\downarrow$ at site ${\bf i}$ for the upper and lower layers, respectively. The spin operator is $\bm{S}_{{\bf i},c}=c_{\bf i}^\dagger \bm{\sigma} c_{\bf i}$, where $c_{\bf i}=(c_{{\bf i}\uparrow},c_{{\bf i}\downarrow})^{\mathrm{T}}$ is a two-component spinor and $\bm{\sigma}$ denotes the vector of Pauli matrices. The occupation number operator is $n_{{\bf i},c}=c_{\bf i}^\dagger c_{\bf i}=\sum_\sigma c_{{\bf i}\sigma}^\dagger c_{{\bf i}\sigma}$.

The noninteracting part of the Hamiltonian, given by the first two lines of Eq.~(\ref{Hamiltonian}), comprises the intralayer nearest-neighbor hopping $t_{\parallel}$, the interlayer on-site hopping $t_{\perp}$, and the chemical potential $\mu$. The interaction terms in the last line are interlayer on-site interactions: the Heisenberg spin-exchange coupling $J_{\perp}$ and the Coulomb repulsion $V$.

The absence of sign problems in our model traces back to a spin-3/2 fermionic system originating from the time-reversal symmetry of Kramers doublets within the SO(5) algebra~\cite{PhysRevLett.91.186402}, which is equivalent to a spin-1/2 bilayer model~\cite{PhysRevB.70.220505,PhysRevB.106.054510,PhysRevB.107.214509,PhysRevB.71.155115}. Leveraging the Kramers invariance, our model can be recast in the form
\begin{eqnarray}
\label{HamiltonianV2}
H&=&-t_{\parallel}\sum_{\langle{\bf ij}\rangle}(\psi_{\bf i}^\dagger \psi_{\bf j}
+\mathrm{H.c.})-t_{\perp}\sum_{{\bf i}}\psi_{\bf i}^\dagger \Gamma^{5} \psi_{\bf i} \nonumber\\
&&-\mu\sum_{{\bf i}}\psi_{\bf i}^\dagger \psi_{\bf i} 
-\sum_{\bf i}\left\{ \frac{g_1}{2}\sum_{ a=1,5}\left(\frac{1}{2}\psi_{\bf i}^\dagger \Gamma^{a} \psi_{\bf i}\right)^2 \nonumber \right.\\
&& \left. +\frac{g_2}{2} \left[ \sum_{a=2,3,4}\left(\frac{1}{2}\psi_{\bf i}^\dagger \Gamma^{a} \psi_{\bf i}\right)^2+3(\psi_{\bf i}^\dagger \psi_{\bf i}-2)^2 \right] \right\},
\end{eqnarray}
with $\psi_{\bf i}=\left[c_{{\bf i}\uparrow},c_{{\bf i}\downarrow},d_{{\bf i}\uparrow},d_{{\bf i}\downarrow}\right]^{\mathrm{T}}$ and the five $\Gamma$ matrices~\cite{PhysRevLett.91.186402}
\begin{equation}
\label{gamma}
\Gamma^1=\begin{pmatrix}0&-iI\\iI&0\end{pmatrix};
\Gamma^{2,3,4}=\begin{pmatrix}\boldsymbol{\sigma}&0\\0&-\boldsymbol{\sigma}\end{pmatrix};
\Gamma^5=\begin{pmatrix}0&I\\I&0\end{pmatrix},
\end{equation} 
where $I$ denotes the $2\times2$ identity matrix and $\boldsymbol{\sigma}$ the vector of Pauli matrices. This Hamiltonian is equivalent to Eq.~(\ref{Hamiltonian}) with the parameter relations $g_1=\frac{1}{4}V+\frac{3}{16}J_{\perp}$ and $g_2=-\frac{1}{4}V+\frac{1}{16}J_{\perp}$.

The PQMC algorithm obtains the ground-state wave function $|\Psi_g\rangle$ by projecting a nondegenerate trial wave function $|\Psi_t\rangle$ ($\langle \Psi_t|\Psi_g\rangle \neq 0$) using the projection operator $e^{-\theta H/2}$. The details are provided in Ref.~\cite{Assaad2008}. In simulations, a potential consequence of the Hubbard-Stratonovich transformation, which decouples four-component interaction terms, is the emergence of negative fermion determinants, leading to the notorious sign problem. Here, owing to Kramers invariance, the fermion matrix produces eigenvalues in the form of complex-conjugate pairs~\cite{PhysRevB.71.155115}, ensuring that its determinant remains positive definite when $g_1$ and $g_2$ are non-negative, i.e., $-\frac{3}{4}J_{\perp}\leq V \leq \frac{1}{4}J_{\perp}$. In this work, we set $t_{\parallel}=1$ as the energy unit, and fix $J_{\perp}=2$ with $t_{\perp}=0.1$ to achieve significant interlayer AFM exchange coupling. We then focus on the effects of interlayer Coulomb repulsion $V$ and hole doping $x$.

To characterize the ILCs, we employ the structure factor of the interlayer currents:
\begin{eqnarray}
\label{uniform}
O_{\textrm{ic}}({\bf q})=\frac{1}{N}\sum_{{\bf i},{\bf j}}e^{i{\bf q}\cdot({\bf i}-{\bf j})}\langle \mathcal{O}^{\textrm{ic}}_{\bf i}\mathcal{O}^{\textrm{ic}}_{\bf j} \rangle,
\end{eqnarray}
where $N$ is the number of unit cells, and the interlayer current order parameter is defined as
\begin{eqnarray}
\label{ic_order}
\mathcal{O}^{\textrm{ic}}_{\bf i}=\frac{i}{2}\sum_{\sigma}(c^{\dagger}_{{\bf i}\sigma}d_{{\bf i}\sigma}-d^{\dagger}_{{\bf i}\sigma}c_{{\bf i}\sigma}).
\end{eqnarray}
The ILCs feature a staggered pattern of interlayer currents accompanied by alternating intralayer currents, as shown in the inset of Fig.~\ref{Fig_1}(b). In this study, we adopt the staggered interlayer current component as the representative descriptor of the global ILC order, given its dominant contribution. Consequently, the ILC order is characterized by the structure factor at wavevector $(\pi,\pi)$, which we define as $O_{\textrm{ILC}}\equiv O_{\textrm{ic}}(\pi,\pi)$. We have also verified the intralayer current pattern; please see {\color{blue}SM, Sec.~S1} for further details. It is worth noting that an interlayer bond order, which can be induced by $J_{\perp}$~\cite{PhysRevB.111.144514}, is degenerate with the interlayer current order in the absence of interlayer hopping $t_{\perp}$, but is suppressed once $t_{\perp}$ is turned on, becoming subdominant to the current order. Therefore, we do not consider the bond order in this work [please see {\color{blue}SM, Sec.~S2} for a detailed discussion].

Given that ILCs are interlayer fluctuations, they are expected to play a role in interlayer pairing. Indeed, it is also confirmed that interlayer pairing is the dominant pairing mechanism in the SC channels of this system [please see {\color{blue}SM, Sec.~S4}]. We therefore define the IS-SC pairing structure factor as:
\begin{eqnarray}
\label{sc}
O_{\textrm{sc}}({\bf q})=\frac{1}{N}\sum_{{\bf i},{\bf j}}e^{i{\bf q}\cdot({\bf i}-{\bf j})}\langle \Delta^{\dagger}_{\textrm{s}}({\bf i})\Delta_{\textrm{s}}({\bf j}) \rangle,
\end{eqnarray}
where the interlayer $s$-wave pairing order parameter is defined as
\begin{eqnarray}
\label{sc_order}
\Delta^{\dagger}_s({\bf i})=\frac{1}{\sqrt{2}}\left(c_{{\bf i}\uparrow}d_{{\bf i}\downarrow}-c_{{\bf i}\downarrow}d_{{\bf i}\uparrow}\right)^{\dagger}.
\end{eqnarray}
The uniform component $O_{\textrm{sc}}(0,0)$ is taken as the measure of IS-SC order, denoted by $O_{\textrm{IS-SC}}$.

To further substantiate our findings, we calculate the correlation length at fixed system size $L$, defined as
\begin{eqnarray}
\label{length}
\xi(L)=\frac{1}{2\sin(\pi/L)}\left( \frac{O_{\alpha}({\bf Q})}{O_{\alpha}({\bf Q}+\delta{\bf q})}-1\right)^{1/2},
\end{eqnarray}
where ${\bf Q}$ denotes the ordering wavevector associated with each order parameter, and $\delta{\bf q}=(2\pi/L,0)$ represents the minimum momentum shift~\cite{PhysRevB.89.094516}. This dimensionless quantity $\xi(L)$ directly captures the competition among different orders.

\noindent
\underline{\it Conflict of interest:}~~
The authors declare that they have no conflict of interest.

\noindent
\underline{\it Acknowledgements:}~~
This work is supported by NSFC (12474218, 12234016 and 12174317 ) and Beijing Natural Science Foundation (No. 1242022 and 1252022). 
This work has also been supported by the New Cornerstone Science Foundation.
The numerical simulations in this work were performed at the HSCC of Beijing Normal University.

\noindent
\underline{\it Author contributions:}~~
Z. Fan performed the simulations, analysed the figures, and wrote the paper; R. Ma and S. Chesi participated in discussion and writing; C. Wu provided academic guidance; T. Ma directed the investigation and writing of the paper. The manuscript reflects the contributions of all authors.

\bibliography{reference}

\newpage
\noindent
\clearpage
\setcounter{figure}{0}\setcounter{table}{0}\setcounter{equation}{0}
\renewcommand{\thefigure}{S\arabic{figure}}
\renewcommand{\thetable}{S\arabic{table}}
\renewcommand{\theequation}{S\arabic{equation}}
\onecolumngrid

\begin{center}{\large\bfseries Supplementary Materials for ``Time-reversal symmetry breaking superconductivity in the presence of loop-current fluctuations''}\end{center}


\noindent
\noindent

This Supplementary Materials provide supplementary numerical evidence and symmetry analyses that validate the key findings of the main text. We characterize the interlayer loop-current pattern and present momentum-space characterizations of the interlayer loop-current and superconducting order parameters, demonstrating their respective staggered and uniform structures. We examine the comparison between interlayer current order and bond order, showing that finite interlayer hopping suppresses the latter and dominates the former. Additional analyses of the spin structure factor reveal the response of antiferromagnetic correlations to interlayer repulsion $V$ and doping $x$, and confirm that antiferromagnetic correlations remain subdominant across all parameter regimes. Finally, we compare the correlation lengths of competing pairing symmetries, establishing the robust dominance of interlayer $s$-wave pairing in the superconducting channel. These results collectively support the phase diagram and the interplay between loop-current fluctuations and superconductivity reported in the main text.

\noindent

\noindent
\subsection{S1. The spontaneous ILC and IS-SC characterized in the Momentum Space}

As sketched in Fig.~1 of the main text, the ILC exhibits two features: staggered interlayer currents between the bilayers and alternating intralayer source-to-drain currents within each layer. We also characterize the latter through an intralayer source-current order parameter, defined as:
\begin{eqnarray}
\label{sourCurrent}
\mathcal{O}^{\textrm{isc}}_{\bf i}=i\sum_{\sigma,{\bf d}}(c_{{\bf i}+{\bf d} \sigma}^\dagger c_{{\bf i} \sigma} - c_{{\bf i} \sigma}^\dagger c_{{\bf i}+{\bf d} \sigma}),
\end{eqnarray}
where the index ${\bf d}={\bf x}, {\bf y}$ denotes the direction vectors to nearest-neighbor sites. This order in the upper layer (described by $c$-operators) is equivalent to that in the lower layer (described by $d$-operators). The structure factor of this order is calculated using the same framework as that of the interlayer current (Eq.~(4) in the main text).

\begin{figure}[htbp]
\includegraphics[scale=0.045]{Fig_S1}
\caption{\label{Fig_S1} Structure factors $O({\bf q})$ for both (a) interlayer current and (b) intralayer source current within \emph{c/d} layer for half-filling and different interlayer repulsion $V=0.0$ and $V=0.3$ at the high-symmetry paths of the Brillouin zone. The maximal points for both (a) and (b) emerges at staggered momentum ($\pi,\pi$), and the corresponding current patterns are attached at (a) and (b), respectively. }
\end{figure}

To identify the ILCs, we measure the structure factors for both interlayer and intralayer source currents, as shown in Fig.~\ref{Fig_S1}. 
For the interlayer component, the staggered interlayer current order is characterized by the wavevector $(\pi,\pi)$, as illustrated in Fig.~\ref{Fig_S1}(a). 
This order is significantly enhanced by $V$. 
For the intralayer component, the structure factor peaks at $(\pi, \pi)$, indicating alternating source-to-drain currents---the intralayer feature of the ILCs---as shown in Fig.~\ref{Fig_S1}(b). 
Together, these interlayer and intralayer current patterns constitute the spontaneous ILC, breaking both time-reversal and translational symmetries.

We find that the intralayer source-to-drain currents are always accompanied by a superior staggered interlayer current order. 
Therefore, we employ the staggered interlayer current order as the representative descriptor of the ILC state, owing to its dominant contribution.

\begin{figure}[htbp]
\includegraphics[scale=0.36]{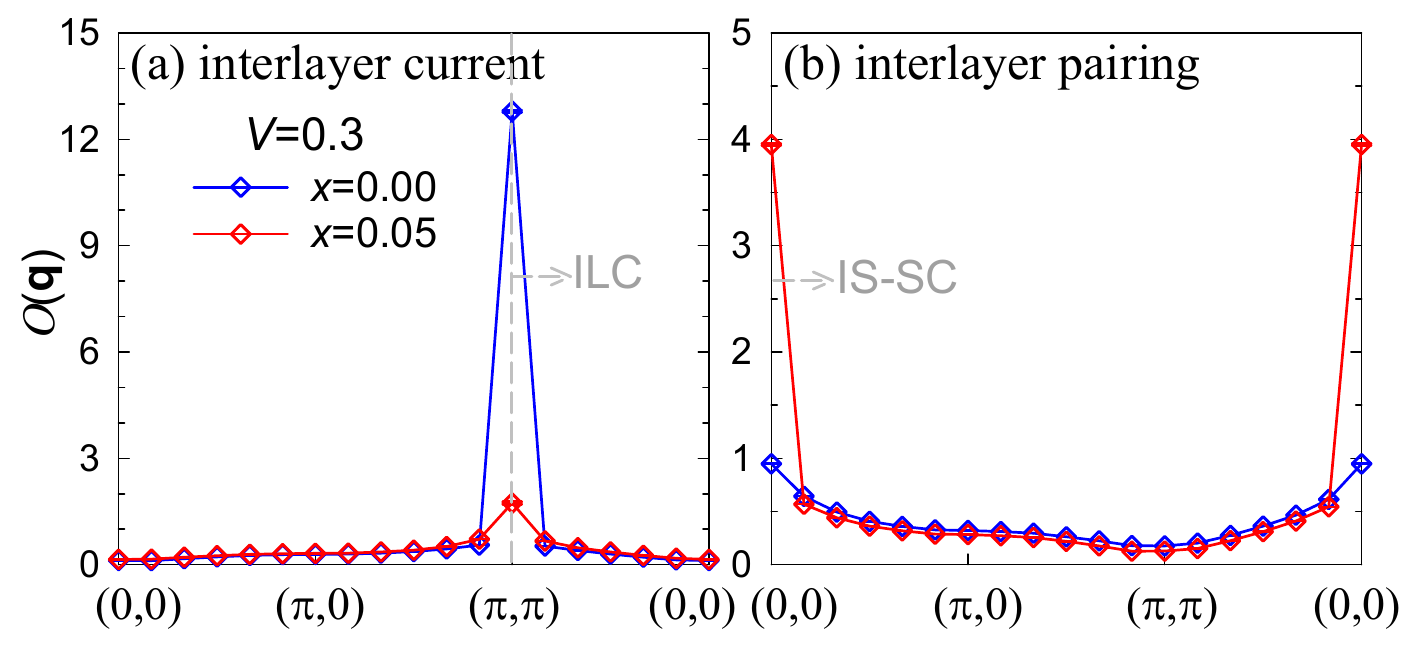}
\caption{\label{Fig_S2} Structure factors $O({\bf q})$ for both (a) interlayer current and (b) interlayer pairing with different doping $x$ for interlayer repulsion $V=0.3$ at the high-symmetry paths of the Brillouin zone. The point at staggered momentum ($\pi,\pi$) at (a) is used to identify the ILCs, and that of uniform momentum ($0,0$) at (b) characterize the IS-SC. }
\end{figure}

Next, we examine the doping evolution of both interlayer current and pairing correlations in momentum space, as shown in Fig.~\ref{Fig_S2}. 
The interlayer current structure factor retains its prominent maximum at $(\pi,\pi)$ upon doping, although its magnitude decreases---indicating that the ILC pattern persists as a characteristic fluctuation. 
For the superconducting component, uniform IS-SC pairing dominates at $(0,0)$ and is enhanced with increasing doping. 
Notably, our model shows no evidence of pairing density waves at finite momentum, in contrast to proposals for kagome systems with loop currents~\cite{zhou2022chern,Li_2025}.

\noindent
\subsection{S2. The interlayer current and bond order introduced by $J_{\perp}$}

In the bilayer framework, electronic hopping between layers can give rise to two distinct interlayer orders: the interlayer current order $\mathcal{O}^{\textrm{ic}}$ associated with imaginary hopping (as described in the main text), and the interlayer bond order $\mathcal{O}^{\textrm{ib}}$ associated with real hopping. These two orders are defined as follows:
\begin{eqnarray}
\label{co}
\mathcal{O}^{\textrm{ic}}_{\bf i}=\frac{1}{2}\sum_{\sigma}(ic^{\dagger}_{{\bf i}\sigma}d_{{\bf i}\sigma} +  \mathrm{H.c.}).
\end{eqnarray}
\begin{eqnarray}
\label{bo}
\mathcal{O}^{\textrm{ib}}_{\bf i}=\frac{1}{2}\sum_{\sigma}(c^{\dagger}_{{\bf i}\sigma}d_{{\bf i}\sigma} + \mathrm{H.c.}).
\end{eqnarray}

The two orders are structurally analogous, corresponding to the imaginary and real parts of the interlayer hopping amplitude, respectively. A recent study on the $t$-$J_{\parallel}$-$J_{\perp}$ model relevant to nickelates indicates that an interlayer bond order phase can be triggered by $J_{\perp}$ with ordering momentum close to $(\pi,\pi)$~\cite{PhysRevB.111.144514}. This suggests a near-degeneracy between the real and imaginary components (i.e., the bond and current orders defined above). We therefore calculate the structure factors of both $\mathcal{O}^{\textrm{ic}}$ and $\mathcal{O}^{\textrm{ib}}$ to examine their relative stability in our model.

\begin{figure}[htbp]
\includegraphics[scale=0.39]{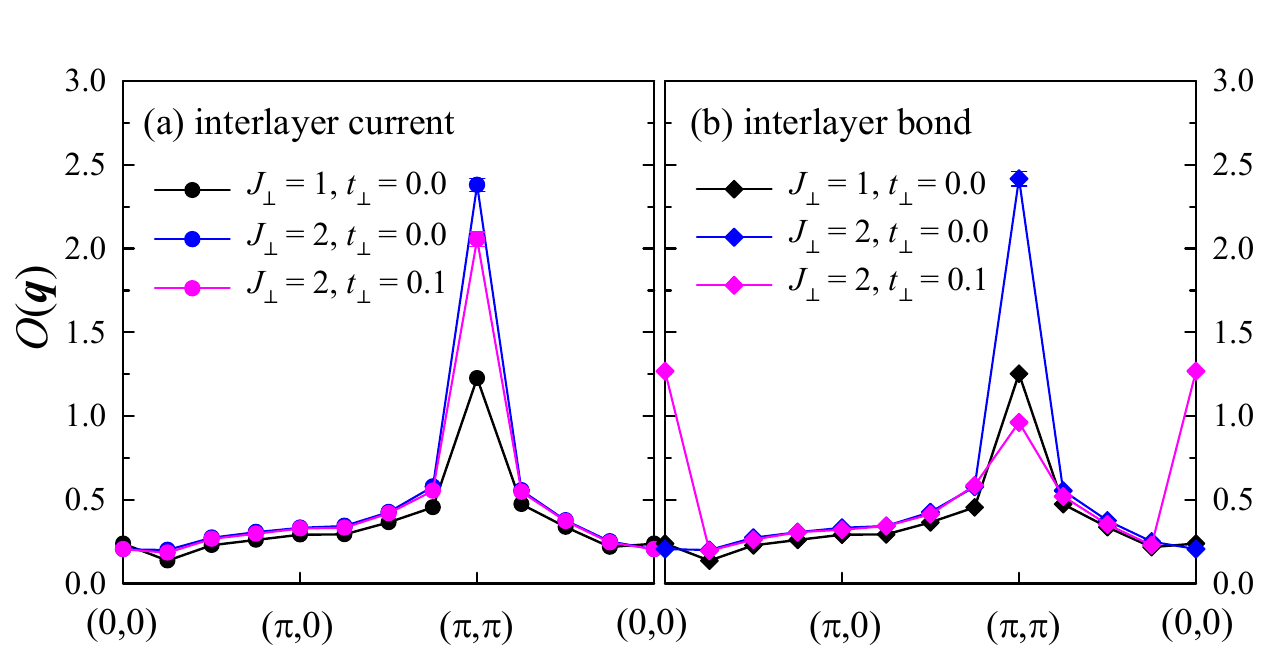}
\caption{\label{Fig_S5} The structure factors of both (a) interlayer current and (b) interlayer bond at different interlayer spin-exchange $J_{\perp}$ and interlayer hopping $t_{\perp}$, at the high-symmetry paths of the Brillouin zone. The two orders are characterized by the momentum $(\pi,\pi)$, which exhibit the high degree of degeneracy and triggered by $J_{\perp}$ at $t_{\perp}=0$ case. While the including of $t_{\perp}$ breaks the degeneracy.}
\end{figure}

As shown in Fig.~\ref{Fig_S5}, we compare systems with different interlayer spin-exchange couplings $J_{\perp}$ and interlayer hopping amplitudes $t_{\perp}$. For $t_{\perp}=0$, the peaks of both the interlayer current and bond structure factors occur at momentum $(\pi,\pi)$ and are induced by $J_{\perp}$ simultaneously. Furthermore, the two structure factors exhibit nearly identical values at all high-symmetry points, reflecting a high degree of degeneracy. This degeneracy is lifted once $t_{\perp}$ is turned on. The current order is suppressed by $t_{\perp}=0.1$, but remains robust at $(\pi,\pi)$. By contrast, the bond order experiences a competition between momentum $(0,0)$ (favored by $t_{\perp}$) and $(\pi,\pi)$ (favored by $J_{\perp}$), resulting in a superior interlayer current order at $(\pi,\pi)$ for $t_{\perp}=0.1$. Therefore, we do not consider the bond order in our work.

\noindent
\subsection{S3. The Spin Structure Factor for the Antiferromagnetic Order}
For completeness, we also examine the spin structure factor to assess possible AFM correlations. The spin structure factor is defined as
\begin{eqnarray}
\label{uniform}
O_{\textrm{s}}({\bf q})=\frac{1}{N}\sum_{{\bf i},{\bf j}}e^{i{\bf q}({\bf i}-{\bf j})}\langle \mathcal{O}^{\textrm{s}}_{\bf i}\mathcal{O}^{\textrm{s}}_{\bf j} \rangle,
\end{eqnarray}
with the interlayer AFM order parameter
\begin{eqnarray}
\label{spin_order}
\mathcal{O}^{\textrm{s}}_{\bf i}=\frac{1}{2}\left(c^{\dagger}_{{\bf i}\uparrow}c_{{\bf i}\uparrow}-c^{\dagger}_{{\bf i}\downarrow}c_{{\bf i}\downarrow}-d^{\dagger}_{{\bf i}\uparrow}d_{{\bf i}\uparrow}+d^{\dagger}_{{\bf i}\downarrow}d_{{\bf i}\downarrow}\right).
\end{eqnarray}
The AFM spin structure factor at $(\pi,\pi)$ is denoted by $O_{\textrm{AFM}}$ below.

\begin{figure}[htbp]
\includegraphics[scale=0.5]{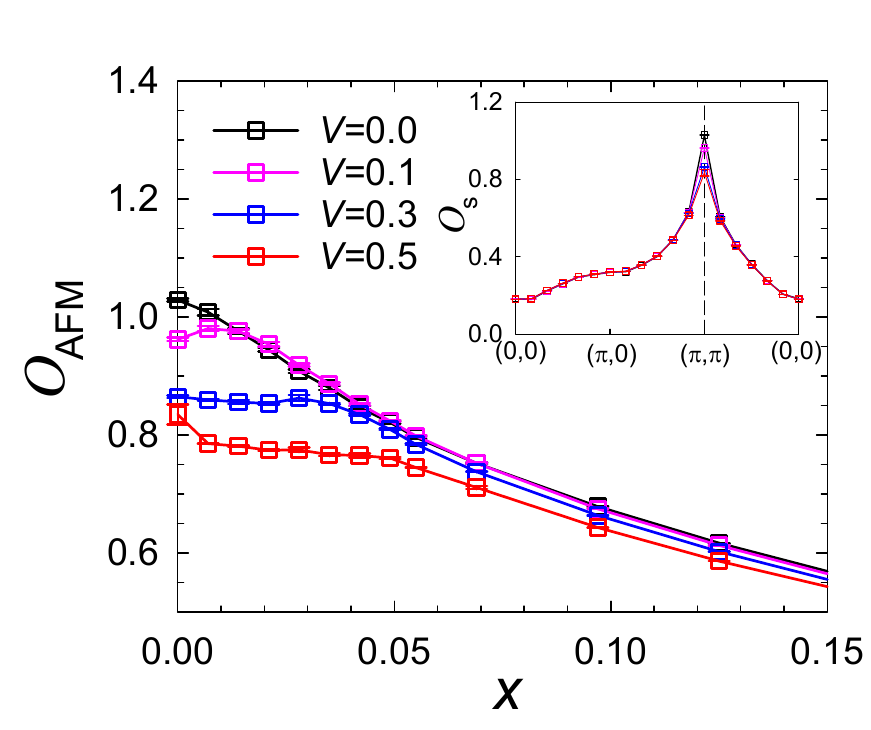}
\caption{\label{Fig_S3}  AFM structure factor $O_{\mathrm{AFM}}$ as functions of doping $x$ with different interlayer repulsion $V$. The inset shows the spin structure factor $O_{\mathrm{s}}$ in the Brillouin zone at half-filling, which is dominated by the AFM order at ${\bf q}=(\pi,\pi)$.}
\end{figure}

We then examine the response of the AFM order to interlayer Coulomb repulsion $V$ and hole doping $x$, as shown in Fig.~\ref{Fig_S3}. It is well established that on-site Coulomb repulsion $U$ in the monolayer Hubbard model stabilizes robust AFM order~\cite{PhysRevB.31.4403}. For comparison, we examine the AFM order in our bilayer model, where the dominant magnetic interaction is the interlayer spin-exchange coupling $J_{\perp}$ rather than on-site Hubbard $U$. Across all parameters considered in this work, we find a pronounced AFM response at $\mathbf{q}=(\pi,\pi)$, as illustrated by scans along high-symmetry paths in momentum space (e.g., the half-filling case shown in the inset of Fig.~\ref{Fig_S3}). The AFM spin structure factor is suppressed by both $V$ and $x$. Near half-filling, $V$ suppresses AFM order by disrupting the charge-occupation balance between the two layers, thereby reducing the local magnetic moments through interlayer charge transfer. Away from half-filling, the suppression is dominated by hole doping $x$, rendering the effect of $V$ negligible. Nevertheless, compared with $O_{\textrm{ILC}}$ and $O_{\textrm{IS-SC}}$, $O_{\mathrm{AFM}}$ remains weak in magnitude, suggesting that AFM order is subdominant. In the main text, we confirm the absence of long-range AFM order through finite-size scaling analysis.

\noindent
\subsection{S4. The dominance of interlayer s-wave pairing in SC channel}

\begin{figure}[htbp]
\includegraphics[scale=0.36]{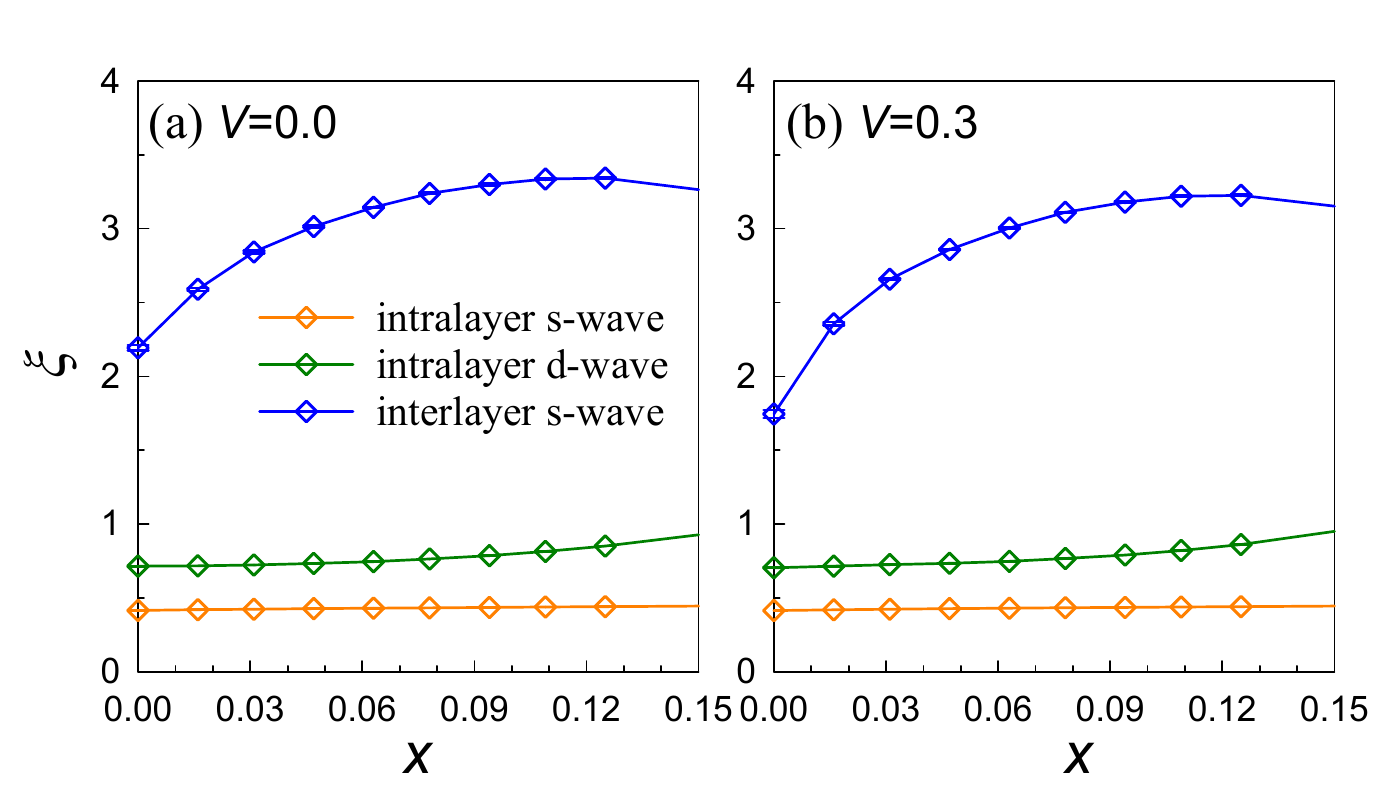}
\caption{\label{Fig_S4} Correlation length $\xi$ as functions of doping $x$ for different pairing symmetries at (a) $V=0.0$ and (b) $V=0.3$. The interlayer $s$-wave pairing is far superior to the interlayer $s$- and $d$-wave pairings in the doping region of this work.}
\end{figure}

The interlayer $s$-wave pairing mechanism has been widely discussed in layered superconducting systems with strong interlayer spin-exchange coupling~\cite{PhysRevB.108.L201108,PhysRevLett.132.146002,PhysRevB.109.045154}. To verify the dominance of interlayer $s$-wave pairing symmetry in our model, we also examine competing intralayer pairing channels. The general intralayer pairing structure factor for the $c$-layer is defined as
\begin{eqnarray}
\label{sc}
O_{\alpha}({\bf q})=\frac{1}{N}\sum_{{\bf i},{\bf j}}e^{i{\bf q}({\bf i}-{\bf j})}\langle \Delta^{\dagger}_{\alpha}({\bf i})\Delta_{\alpha}({\bf j}) \rangle,
\end{eqnarray}
with the intralayer pairing order parameter
\begin{eqnarray}
\label{layer-wave}
\Delta_{\alpha}^{\dagger}({\bf i})=\sum_{\boldsymbol{\delta}_l}f_{\alpha}^{\ast}(\boldsymbol{\delta}_l)\left(c_{{\bf i},\uparrow}c_{{\bf i}+\boldsymbol{\delta}_l,\downarrow}
-c_{{\bf i},\downarrow}c_{{\bf i}+\boldsymbol{\delta}_l,\uparrow}\right)^{\dagger},
\end{eqnarray}
where $\alpha$ represents the corresponding pairing symmetry and $\boldsymbol{\delta}_l$ contributes to the four nearest neighbour site vectors for each site. The intralayer $s$-wave symmetry and $d$-wave symmetry are undertaken by the coefficient $f_{intra-s}(\boldsymbol{\delta}_l)=1$ for $l=1,2,3,4$ and $f_{intra-d}(\boldsymbol{\delta}_l)=+1 (-1)$ for $l=1,3 (2,4)$, respectively.

We directly compare the correlation lengths $\xi$ (Eq.~(8) in the main text) of intralayer $s$-wave, intralayer $d$-wave, and interlayer $s$-wave pairing symmetries through the analysis of the uniform pairing structure factor, as shown in Fig.~\ref{Fig_S4}. The intralayer $s$-wave and $d$-wave pairing correlations remain weak and largely independent of doping $x$ and interlayer Coulomb repulsion $V$ within our parameter range, exhibiting small correlation lengths. For both $V=0$ and $V=0.3$, the correlation length of interlayer $s$-wave pairing far exceeds those of the intralayer channels, demonstrating the robust dominance of interlayer $s$-wave pairing in our model.

\end{document}